 \documentstyle[prl,aps,multicol,psfig]{revtex}
\newcommand{\be}{\begin{equation}}
\newcommand{\ee}{\end{equation}}
\newcommand{\br}{\begin{eqnarray}}
\newcommand{\er}{\end{eqnarray}}

\newcommand{\bd}{\begin{displaymath}}
\newcommand{\ed}{\end{displaymath}}

\newcommand{\bfig}{\begin{figure}}
\newcommand{\efig}{\end{figure}}

\def\lb#1{\label{#1}}
\def\3cdot{\cdot \cdot \cdot}

\def\om0{\omega _0}
\def\Om0{\Omega _0}

\def\rg{\rangle}

\def\text#1{{\rm{#1}}}

\def\->{\rightarrow}
\def\=>{\Rightarrow}
\def\-->{\longrightarrow}
\def\==>{\Longrightarrow}

\def\lbk{\left[}
\def\rbk{\right]}

\def\ox{\otimes}

\def\pr{^\prime}
\def\pr2{^{\prime\prime}}

\def\bfig{\begin{figure}}
\def\efig{\end{figure}}

\begin{document}
 \draft
 \title{Continuous Measurement of Atom-Number
  Moments of a Bose-Einstein Condensate  by Photodetection}
 \author{G. A. Prataviera$^1$\footnote{gap@df.ufscar.br} and M. C. de Oliveira$^2$\footnote{marcos@ifi.unicamp.br} }
 \address{$^1$ Departamento de F\'\i sica, CCET, Universidade Federal de S\~ao Carlos,\\
 Via Washington Luiz Km 235, S\~ao Carlos, 13565-905, SP, Brazil.\\
 $^2$ Instituto de F\'\i sica ``Gleb Wataghin'',  Universidade Estadual de Campinas,\\
 13083-970, Campinas - SP, Brazil.}
 \date{\today}
 \maketitle

 \begin{abstract}
We propose a measurement scheme that allows determination
 of even-moments of a Bose-Einstein condensate (BEC) atom number, in a ring
 cavity, by continuous photodetection of an off-resonant quantized optical
 field.
A fast cavity photocounting process limits the heating of atomic
samples with a relatively small number of atoms, being convenient
for BECs on a microchip scale applications.
 The measurement back-action introduces a
counting-conditioned phase damping, suppressing the condensate
typical collapse and revival dynamics.
 \end{abstract}
\begin{multicols}{2}
The recent achievement of Bose-Einstein Condensates (BECs) trapped
near the surface of magnetic microchip traps
 \cite{chip} has led to a new promising system for the
 development of emerging technologies based on BECs,
such as trapped-atom interferometry \cite{interferometry} or
atom-based quantum information processing (QUIP) \cite{qc1}, due
the high degree of control achieved over the atomic sample. A
fundamental issue for implementing those technologies on a chip
scale though is the achievement of a non-destructive measurement
of the BEC properties. Particularly, QUIP calls for high precision
non-destructive detection of the BEC atom number \cite{qc1}, which
has proven to be a hard task, attracting considerable attention
\cite{long,hope,horak}.

Since the very early experiments with diluted trapped neutral
atoms \cite{1}, the BEC dynamics monitoring has been achieved
either by absorption or dispersive imaging \cite{stamper-kurn}.
Absorption imaging has the counter-effect of heating up the
condensate, precluding it for latter usage (destructive regime).
On the other hand in dispersive imaging the small phase-shift
suffered by the far-detuned probe light is compensated by a high
intensity. Residual incoherent Rayleigh scattering heats up the
atomic sample through spontaneous emission atomic recoil,
preventing a non-destructive regime as well \cite{stamper-kurn}
for the reduced number of atoms in microchip BECs ($\sim 10^4$)
\cite{chip}. Thus, it is certainly worthwhile to propose
alternative schemes of atom detection that besides being
non-destructive to some extent, could also be useful for feedback
and control of the condensate - a valuable resource for QUIP.

In this Rapid Communication we investigate the information
extracted about a BEC atom number through probe-field continuous
photodetection. Previous treatments on BEC continuous measurements
have been described in Refs. \cite{ruostekoski,corney,guo,dalvit},
differing considerably from our approach and goals. We consider a
BEC trapped inside a ring-cavity fed by two resonant (orthogonaly
polarized) propagating fields - an undepleted probe and a weak
quantum probe field (Fig.1). The presence of the undepleted pump
field allows that the moments of the detected probe field photon
number give direct information about even-moments of the BEC atom
number. Moreover, since the condensate atom number information is
carried by the probe field photocounting statistics, there is no
need for a strong probe field, avoiding thus heating during the
measurement process. Finally, we discuss how the detection
back-action induces phase uncertainty to the condensate state,
suppressing its original collapse and revival dynamics.

 The system, depicted in Fig.1, consists of a
Schr\"{o}dinger field of bosonic two-level atoms with transition
frequency $\nu _{0}$ interacting via electric-dipole with the two
single-mode orthogonally polarized ring-cavity probe and pump
fields of frequencies $\nu _{1}$ and $\nu _{2}$, respectively,
both being far-off resonant from any electronic transition
(Calculation details given in \cite{olivprat}). The eigenstates
for the atoms  are denoted by $\left| k\right\rangle $ with
eigenfrequencies $\omega _{k}$, whose values are dependent on the
trapping conditions. For an atomic cloud well localized both
longitudinally and transversally relative to the cavity round-trip
($L$) and to the cavity field beam waist ($S$), respectively, the
field can be assumed uniform in its vicinity, such that the
coupling between atoms and pump and probe fields is approximately
constant. In the far-off resonance regime the $k$-excited state
population is negligible, and the collision between excited atoms,
and between the excited and the ground state atoms can be
neglected. In a rotating referential with the frequency $\nu _{2}$
the Hamiltonian writes
\begin{eqnarray}
H &=&\hbar \sum_{k}\left[ \omega _{k}c_{k}^{\dagger }c_{k}+\Delta
_{k}a_{k}^{\dagger }a_{k}\right] +{\hbar }\sum_{jklm}\kappa
_{jklm}c_{j}^{\dagger }c_{k}^{\dagger }c_{l}c_{m}
\nonumber \\
&&+\hbar \sum_{n=1}^2\sum_{jk}\left( g_{n}\left\langle j\right|
e^{i{\bf k}_{n}\cdot {\bf r}}\left| k\right\rangle
b_na_{j}^{\dagger }c_{k}+h.c.\right)\nonumber\\
&&+\hbar \delta b_1^{\dagger }b_1+\hbar\left(F b_2^\dagger+F^*
b_2\right), \label{m1}
\end{eqnarray}
where $\Delta _{k}=\omega _{k}+\Delta$ and $\delta =\nu _{2}-\nu
_{1}$, being $\Delta=\nu_{0}-\nu _{2}$ is the detuning between
pump and atom. $c_k$ and $a_k$ are the annihilation operators for
atoms with $k$ in the ground and excited state, respectively, and
$\kappa_{jklm}$ is the collision strength between ground state
atoms. The third term in the RHS of Eq. (\ref{m1}) is the
interaction between the atoms and the probe ($b_1$) and pump
($b_2$) fields (with coupling constants $g_1$ and $g_2$,
respectively), whose wave vectors ${\bf k_{1(2)}}$ must satisfy
 $|{\bf k_{1(2)}}|=2\pi
 n/L$, with $n$ integer.  The in-cavity probe field is related to the
 input field ($b_1^{in}$) by $b_1=\sqrt{T_0}\,b_1^{in}$ (neglecting fluctuations),
  where $T_0$ is the mirror 0 transmission index.
 The field $b_2$ external pumping is given by the last term of (\ref{m1}), where $|F|$
is the external resonant driving field strength.  If the pump beam
cavity-loss is considerably higher than the coupling constants the
pump average photon number can be kept constant (undepleted), due
to the pump-loss competition. This assumption allows the pump
field to be treated as a c-number, and also avoids the atomic
sample heating through residual incoherent Rayleigh scattering by
setting a low steady pump intensity. Since we also require that
the probe field loss rate is smaller than the photocounting rate,
the pump is set to a $\parallel$-polarization (to the table top)
while the probe is set to a $\perp$-polarization. The cavity
mirror $1$ thus must have distinct reflection indexes
$R_1^\perp\gg R_1^\parallel$. Assuming the bad-cavity limit  for
the $\parallel$-polarization ($\gamma_{inc}^\parallel\gg
\frac{|g_1|^2}{\gamma_{inc}^\parallel},\frac{|g_2|^2}{\gamma_{inc}^\parallel}\gg\Gamma$),
with
 $\gamma_{inc}^\parallel\propto T_1^\parallel=1-R_1^\parallel$ and
 $\Gamma$ the atomic spontaneous emission rate, the pump field can be adiabatically
 eliminated such that $b_2$ can be replaced by $-i
 F/\gamma_{inc}^\parallel$. Remark that the probability of atomic spontaneous emission
($P_e$) is also reduced inside resonators \cite{long,zimmermann}
with high finesse ${\cal F}$, since the {\it per photon}
probability of spontaneous emission goes with $P_e\propto{\cal
F}$, and the required number of the probe beam photons for
reliable detection is $\bar{N}\propto{\cal F}^{-2}$, thus the
total number of spontaneous scattering events is $\bar{N}
P_e\propto{\cal F}^{-1}$ \cite{long}.

In the limit of large detuning $|g_i/\Delta|\ll 1,\;\; i=1,2$ and
$\omega _{k}/\Delta\ll 1$, $\Delta _{k}\approx\Delta$
\cite{javanainen2}. Thus atomic spontaneous emission can be
neglected and the excited states operators $a_k$ are eliminated
adiabatically resulting in the following effective hamiltonian
\begin{eqnarray}
H_{eff} &=&\hbar \delta b^{\dagger }b +\hbar \sum_{k}(\omega
_{k}+\frac{|\widetilde{g_{2}}|^2}{\Delta})c_{k}^{\dagger }c_{k}\nonumber \\
&&+\hbar \sum_{klmn}{\kappa_{klmn}}c_{k}^{\dagger }c_{l}^{\dagger
}c_{m}c_{n}+\hbar\frac{|g_{1}|^2}{\Delta} b^{\dagger }b
\sum_{k}c_{k}^{\dagger
}c_{k}\nonumber\\
&&+\hbar \sum_{kl}\left(\frac{ g_{1}^\ast \widetilde{g_2}}{\Delta}
\langle k|e^{-i({\bf k}_1-{\bf k}_2)\cdot {\bf r}}|l\rangle
b^{\dagger
}\right.\nonumber\\
&&\left. +\frac{ g_{1} \widetilde{g_2^\ast}}{\Delta}\langle
k|e^{i({\bf k}_1-{\bf k}_2)\cdot {\bf r}}|l\rangle b
\right)c_k^\dagger c_l,\label{eff1}
\end{eqnarray} where $\widetilde{g_2}\equiv -i g_2
F/\gamma_{inc}^\parallel$ is the effective coupling, and we have
defined $b_1\equiv b$. Hamiltonian (\ref{eff1}) is the prototype
for atom-optic parametric amplification \cite{meystre}, where
atoms in the ground state are transferred to side modes states.
However, we are interested in the situation where no optical
inter-mode excitation occurs. In the ring-cavity arrangement ${\bf
k_{1(2)}}$ (with $|{\bf k_{1(2)}}|=2\pi n/L$) are both co-linear
to the longitudinal dimension of the condensate $L_c$, which is
taken to be very small compared to the cavity round-trip length
$L$. Thus $\langle k|e^{i({\bf k}_1-{\bf k}_2)\cdot {\bf
r}}|l\rangle\approx \delta_{k,l}$ whenever $2\pi nL_c/L\rightarrow
0$, and no inter-mode excitation occurs. This embodies the
specific case of ${\bf k_1\approx k_2}$ (and thus $\delta=0$),
which we consider hereafter.
To simplify we further assume a pure condensate with all atoms in
the $c_0$ mode, the hamiltonian finally reduces to
\begin{eqnarray}\label{hamil}
H_{eff} &=&\hbar (\omega
_{0}+\frac{|\widetilde{g_{2}}|^2}{\Delta})c_{0}^{\dagger
}c_{0}+\hbar
{\kappa}c_{0}^{\dagger }c_{0}^{\dagger }c_{0}c_{0}\nonumber\\
&&+\hbar\frac{|g_{1}|^2}{\Delta} b^{\dagger }b c_{0}^{\dagger
}c_{0}+\hbar \left( \frac{g_1^*\widetilde{g_2}}{\Delta}b^{\dagger
}+ \frac{g_1\widetilde{g_2^*}}{\Delta}b \right)c_0^\dagger c_0.
\end{eqnarray}

 In (\ref{hamil}) we identify two regimes in the
interplay between the pump and probe fields strength: (i) Whenever
$|\widetilde{g_2}/g_1|\ll 1$ the strongest contribution is from
the quantum probe field, including the situation without the
classical pump-field; (ii) otherwise the classical pump field has
an important contribution to the effective hamiltonian. Remark
from (\ref{hamil}) that the condensate atom number
$n_0=c_0^\dagger c_0$ is a non-demolition variable. By varying
$|F/\gamma_\parallel|$ and thus $|\widetilde{g_2}|$ distinct
regimes of quantum nondemolition couplings
\cite{grangier,haroche,viola} are attained. For
$|\widetilde{g_2}/g_1|\ll 1$ the nondemolition regime corresponds
to that considered in Refs. \cite{corney,guo} for BECs atom number
nondemolition measurement, while for $|\widetilde{g_2}/g_1|
\gtrsim1$ features similar to the photon number nondemolition
measurements discussed in Ref. \cite{mi1} are added.
%
%

Now we turn to the photodetection process. To simplify the
photocounting modelling \cite{davies} we firstly assume that no
other incoherent process, such as $\perp$-polarized photon losses,
considerably affects the the probe field dynamics over the
counting time interval. That means to assume
$\gamma_{inc}^\perp\ll\gamma$, where $\gamma$ is the effective
cavity photodetection rate given by $\gamma\approx T_2^\perp
\eta$, where $T_2^\perp$ is the mirror 2 transmission coefficient
and $\eta$ is the output field photodetection rate, neglecting
output field fluctuations \cite{olivprat,milburnli}. The counting
of $k$ photons from the probe field in a time interval $t$ can be
characterized by the linear operation $N_t(k)$ \cite{davies},
acting on the state of the system as $ \rho^{(k)}(t)=N_t(k)\rho
(0)/{\rm Tr}\lbk N_t(k)\rho(0) \rbk$
where $\rho (0)$ is the joint state of the condensate and the
probe field prior turning on the counting process, with
probability $P(k,t)={\rm Tr} [N_t(k)\rho(0)]$. The operation
$N_t(k)$ is written as
\br \label{Nt}%
 N_t(k)&=&\int_0^tdt_k\int_0^{t_k}
dt_{k-1}\cdot\cdot\cdot\int_0^{t_2}
dt_1\nonumber\\
&&\times S_{t-t_k}JS_{t_k-t_{k-1}}\cdot\cdot\cdot JS_{t_1}\, ,
\er%
where $ S_t\rho=e^{Yt}\rho e^{Y^\dagger t}$, with $ Y=-\frac
i\hbar H-R/2$. $H$ is the system Hamiltonian, and $R=\gamma
b^\dagger b$ is the counting rate operator. As such
$J\rho\equiv\gamma b\rho b^\dagger$ stands for the change of the
probe field due to the loss of one counted photon, while $S_t$ is
responsible for the state evolution between counts.

From Eq. (\ref{hamil}) $Y$ becomes
\br \label{Y}
Y&=&-i\left(\omega_0-\kappa+\frac{\widetilde{g_2}^2}{\Delta}\right)n_0-i
\kappa n_0^2\nonumber\\
&&-i(\delta_{n_0}-i\frac\gamma 2)b^\dagger b-i\left(F_{n_0}^* b
+F_{n_0}b^\dagger \right)\er
where we defined $
\label{delta}\delta_{n_0}\equiv\frac{|g_1|^2}{\Delta}n_0,\;\;
F_{n_0}\equiv\frac{g_1^*\widetilde{g_2}}{\Delta} {n_0}$.
We express the
 $N_t (k)$ acting on the joint initial state $\sum_{m}C_{m} |m\rg  \ox |\beta\rg $, where the first ket stands
for the condensate state while the second is the probe-field
state, hereafter assumed as coherent.

After  $k$-count events on the probe field, the conditioned joint
state becomes
\br%
 \rho^{(k)}(t) &=& \frac{1}{k! P(k,t)}
 \sum_{m,m'}C_{m}C^*_{m'} \;\; {\cal F}_{m,m'}^k(t)
 \nonumber\\
&& \times \;e^{\Phi_m(t)+\Phi_{m'}^*(t)}| m\rangle \langle
m'|\otimes| \beta_m(t) \rangle \langle\beta_{m'}(t)|\lb{post},
\er%
where %
\br%
{\cal F}_{m,m'}(t)&\equiv& {\gamma}\left\{-\frac{\Lambda_m
\Lambda_{m'}^*}{\Gamma_m+\Gamma_{m'}^*}\left[e^{-(\Gamma_m+\Gamma_{m'}^*)t}-1\right]\right.\nonumber\\
&&\left. +G_m G_{m'}^* t+i\left[\frac{G_m
\Lambda_{m'}^*}{\Gamma_{m'}^*}
\left(e^{-\Gamma_{m'}^*t}-1\right)\right.\right.\nonumber\\
&&\left.\left.-\frac{G_{m'}^* \Lambda_{m}}{\Gamma_m}
\left(e^{-\Gamma_{m}t}-1\right)\right] \right\},
 \er
 with $\Gamma_m=(i\delta_m+\gamma/2)$, $G_m=F_m/\Gamma_m$, and
 $\Lambda_m=\beta+i G_m$,
for  $\delta_m=\frac{|g_1|^2}{\Delta} m$ and $F_m = \frac
{g_1^*\widetilde{g_2}}{\Delta} m$. In Eq. (\ref{post}),
$\beta_m(t)\equiv
 \Lambda_m e^{-\Gamma_m t}-iG_m
 $ is the label for the probe field coherent state,
 \br\label{phi}
 \Phi_m(t)&\equiv&-\frac 1 2\left(|\beta|^2-|\beta_m(t)|^2\right)+i\left[G_m\Lambda_m\left(e^{-\Gamma_m t}-1\right)\right.\nonumber\\
 &&\left.+\left(i|G_m|^2\Gamma_m^*-\theta_m\right) t\right],\er and
$\theta_m\equiv\left[\omega_0+\frac{|g_2|^2}{\Delta}+\kappa
(m-1)\right]m$ is a phase introduced by the atomic collision
process and the classical pump. The last two terms of $\Phi_m(t)$,
Eq. (\ref{phi}), besides a direct collision process also include
the terms $|G_m|^2$ and $G_m\Lambda_m\left(e^{-\Gamma_m
t}-1\right)$, which are originated by the pump-field, inducing a
collision-like behavior, with diffusion of the condensate state
phase.

The probability to count $k$ photons during the time interval $t$
is given by
\be%
P(k,t) = \frac1 {k!}\sum_{m}|C_{m}|^2{\cal F}_{m,m}^k(t)e^{-{\cal
F}_{m,m}(t)}\lb{probk2}.
\ee%
In regime (i), $\widetilde{g_2}/g_1\ll 1$, the counting
probability (\ref{probk2}) reduces to the Poisson distribution 
 \be
 P(k,t)=
\frac1{k!}[|\beta|^2(1-e^{-\gamma t})]^k
e^{-|\beta|^2(1-e^{-\gamma t})}, \ee independently of the
condensate state and the atom-field coupling as well. The
$r$-moments of $P(k,t)$ for this regime are $
\overline{{k^r}}=\left[|\beta|^2(1-e^{-\gamma t})\right]^r$, and
simply relate to the probe amplitude. However in regime (ii), Eq.
(\ref{probk2}) must be fully considered, and the condensate state
is relevant for the photocounting probability distribution. Thus
inference about the condensate atom number moments can be given by
the photocounting distribution. The $r$-moments of (\ref{probk2})
are \br \overline{{k^r}}&=&\sum_m |C_m|^2 {\cal
F}^r_{m,m}(t)=\langle{\cal F}^r_{n_0,n_0}(t)\rangle, \er which in
the long time limit ($\gamma t\gg 1$) goes to
\br \label{mom}\overline{{k^r}}&\approx& (\gamma t)^r
\left\langle\left(\frac{|\frac{g_1}{\Delta}|^2 n^2_0}
{\left|\frac{\gamma}{2\widetilde{g_2} }\right|^2
 +\left|\frac{g_1}{\Delta}\right|^2 \left|\frac{g_1}{\widetilde{g_2} }\right|^2
 n^2_0}\right)^r\right\rangle.\er
The limit
${\frac{\gamma}{2\Delta}}\gg\left|\frac{g_1}{\Delta}\right|^2$
gives the central result of this paper, since we may approximate
(\ref{mom}) by
 \br \label{mom2}\overline{{k^r}}&\approx& (\gamma t)^r
\left|\frac{2g_1\widetilde{g_2}}{\gamma\Delta}\right|^{2r}\left\langle
 n^{2r}_0\right\rangle,\er and the even-moments of the condensate
 atom-number are directly given by the moments of the number of photocounts.
 Particularly, for a BEC in a Fock state $\sqrt{\overline k}$ gives
 a null-uncertainty measure of  the condensate $\left\langle
 n_0\right\rangle$.

In the opposite limit,
${\frac{\gamma}{2\Delta}}\ll|\frac{g_1}{\Delta}|^2$, the
photocounting moments give
 \br \label{mom2}\overline{{k^r}}&\approx& (\gamma t)^r
 \left|\frac{\widetilde{g_2}}{g_1}\right|^{2r},\er
and thus the fields strength ratio is dynamically probed {\it in
situ}, while the condensate is inside the cavity,
 by the determination of the average number of counted photons at the slow rate
 ${\frac{\gamma}{2\Delta}}\ll|\frac{g_1}{\Delta}|^2$.

The important time scale parameter for determination of the
condensate atom number even-moments by photocounting is the
effective photocounting rate $\gamma$. Since the undepleted
classical pump field approximation is valid only in the
$\parallel$-polarization
 bad-cavity limit ($\gamma_{inc}^\parallel\gg \frac{|g_1|^2}{\gamma_{inc}^\parallel},
 \frac{|g_2|^2}{\gamma_{inc}^\parallel}$) we must also have
  $\gamma_{inc}^\parallel\gg\gamma$. The ability to build-up a ring cavity with high finesse at
the microchip surface could represent a restriction, but recent
efforts have been made in the study of properties of ultra cold
atomic samples inside a ring cavity, which could attain finesses
as high as 170000 \cite{zimmermann}. In fact, a high finesse
cavity is necessary only when the small phase shift has to be
compensated by a large intensity field, such as in dispersive
imaging, since information about the BEC is carried by the probe
field phase. However, in our proposal the pump and probe
intra-cavity fields can be both set at low intensity, which limits
the effects of incoherent Rayleigh scattering through spontaneous
emission during the photocounting period. If every atomic
spontaneous emission heats the condensate in about an atomic
recoil energy $E_R$, we can estimate the total heating due the
fraction $N_e=P_e \langle
 n_0\rangle$ of atoms suffering spontaneous emission, where $P_e\propto\Gamma/\Delta^2$ is
the per photon spontaneous emission probability in the far-off
resonance regime with the intra-cavity spontaneous emission rate
$\Gamma$. The BEC heating due the interaction with the probe light
with $I=\langle b^\dagger b\rangle$ photons amounts to $\Delta
T\propto 2E_RI \langle
 n_0\rangle\Gamma/3k_B\Delta^2$. For the regime of
 ${\frac{\gamma}{2\Delta}}\gg\left|\frac{g_1}{\Delta}\right|^2$ of
optimal detection of the atom number moments we can set a limiting
$\Gamma$ such that the heating is negligible, {\it i. e}, by
considering
$\Gamma\ll\frac{|g_1|^2}{\gamma_{inc}^\parallel},\frac{|g_2|^2}{\gamma_{inc}^\parallel}\ll\gamma\ll\gamma_{inc}^\parallel$.
 Since $\gamma=\eta T_2^\perp$, the above limit can be
conveniently reached with a high $\perp$-transmission coefficient
mirror and a reasonably fast photodetector.

 Despite  the heating process being negligible there will always be a back-action on the condensate
state due the continuous measurement process. Only if the
condensate is initially in a Fock state, an eigenstate of the
nondemolition variable, is that the condensate will evolve freely
independently of the counting probability. The same is valid for
the diagonal elements of the unconditioned state,
\br%
\rho_c(t) &=& \sum_k P(k,t)\rho^{(k)}_c(t)\nonumber\\&=&
\sum_{m,m'}C_{m}C^*_{m'} e^{\Phi_m(t)+\Phi_{m'}^*(t)+{\cal
F}_{m,m'}(t)}\nonumber\\
&&\langle\beta_{m'}(t)| \beta_m(t) \rangle| m\rangle \langle m'|,
\lb{pre}
\er%
since $\langle m|\rho_c(t)|m\rangle=|C_m|^2$. The off-diagonal
elements of  Eq.
(\ref{pre}) are evidences of back-action over the condensate state
phase. Obviously this implies that a condensate in a completely
mixed state will not suffer the back-action effects. Any other
condensate state will be affected by the collision-like terms
$|G_m|^2$ and $G_m\Lambda_m\left(e^{-\Gamma_m t}-1\right)$
   from  Eq. (\ref{phi}).
  The counting process induces an irreversible phase-damping,
  inhibiting  the well known coherent collapse
     and revival dynamics of the condensate state \cite{parkins}.
      Interestingly though, the $k$-counts conditioned phase damping
     does not appear when no photons are
       detected, $k=0$, and  the BEC state evolves with 
       its typical collapse and revival dynamics.

 In conclusion, we have
investigated the measurement over the BEC inside a ring cavity
that can be achieved through continuous photodetection of a
quantum probe field. Even-moments of the condensate atom number
can be inferred by the probe field photodetection probability
distribution whenever the photodetector counting rate follows
${\frac{\gamma}{2\Delta}}\gg|\frac{g_1}{\Delta}|^2$. Also if those
rates are higher than the atomic spontaneous emission rate the
condensate heating will be prevented. Although atom number is a
QND variable, there is a back action on the condensate state due
to the counting process, inducing phase-damping over the
condensate state whenever photons are counted. The strong
dependence of the photocounting probability distribution with the
BEC original state suggests that this measurement scheme can be a
useful resource for feedback and control of atomic samples.
Further investigation on those issues for monitoring of
cross-correlation between atoms and light fields together with
calculations on signal to noise ratio, as well as a measurement
resource for atom based quantum information processing will be
addressed elsewhere \cite{olivprat}.

As far as we know, it is still unknown whether surface
interactions reinforced by the cavity will introduce noise
limiting
 the detection process. Besides technical problems
yet to be solved for cavity quantum electrodynamics implementation
on microchips  \cite{long}, we believe that the above proposal
could be implemented, in principle, due the rapid advance on
experimental research.

{We thank G. J. Milburn for his encouragement, and J. Reichel and
to C. Zimmermann for their kind assistance on experimental issues.
This work was supported by FAPESP-Brazil.}
 %
 
\end{multicols}

\begin{figure}
 \centerline{$\;$\hskip 0truecm\psfig{figure=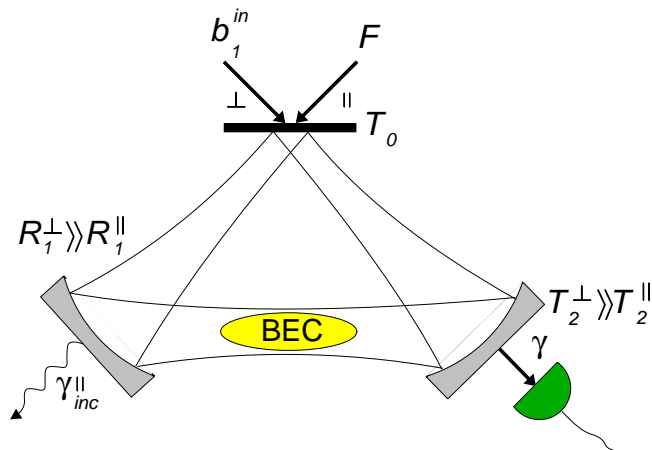,height=6cm}}
\centerline{\caption{ BEC in a ring cavity setup. The pump ($F$)
and probe ($b_1^in$) input fields are $\parallel$ and
$\perp$-polarized, respectively. Mirror 1 and 2
 reflectivities are polarization selective, in order that the in-cavity pump probe is heavily
 damped at mirror 1, while the transmissivity at mirror 2 allows
 that BEC properties be determined by the probe field photocounting at the mirror 2
 output.}}
\end{figure}
\end{document}